\documentclass[11pt,a4paper,reqno]{amsart} 
\usepackage{tikz}
\usetikzlibrary{positioning,decorations.pathmorphing}
\usepackage{amssymb,graphicx}

\usepackage{amssymb}

\newcommand{\koniec}{\begin{flushright}  $\Box $ \end{flushright}}
\newtheorem{theo}{Theorem}[section] 
\newtheorem{prop}[theo]{Proposition}

\newtheorem{defi}[theo]{Definition}
\newtheorem{col}[theo]{Corollary}

\newcounter{mnotecount}[section]

\renewcommand{\themnotecount}{\thesection.\arabic{mnotecount}}

\newcommand{\mnote}[1]
{\protect{\stepcounter{mnotecount}}$^{\mbox{\footnotesize
$
\bullet$\themnotecount}}$ \marginpar{
\raggedright\tiny\em
$\!\!\!\!\!\!\,\bullet$\themnotecount: #1} }

\newcommand{\CP}{\mathbb{CP}}

\newcommand{\R}{\mathbb{R}}

\def\ov{\overline}

\def\be{\begin{equation}}

\def\ee{\end{equation}}

\def\bea{\begin{eqnarray}}
\def\eea{\end{eqnarray}}

\def\ov{\overline}

\newcommand{\bP}{{\bf P}}
\newcommand{\bX}{{\bf X}}
\newcommand{\bY}{{\bf Y}}
\newcommand{\bI}{{\bf I}}
\newcommand{\bT}{{\bf T}}
\newcommand{\bK}{{\bf K}}
\newcommand{\bO}{{\bf O}}
\newcommand{\bDelta}{{\bf \Delta}}

\begin{document}\date{1 January 2020}
\vspace*{-1.0cm}
\title{Conformal and isometric embeddings of gravitational
instantons}
\author{Maciej Dunajski}
\address{Department of Applied Mathematics and Theoretical Physics\\ 
University of Cambridge\\ Wilberforce Road, Cambridge CB3 0WA, UK.}
\email{m.dunajski@damtp.cam.ac.uk}

\author{Paul Tod}
\address{The Mathematical Institute\\
Oxford University\\
Woodstock Road, Oxford OX2 6GG\\ UK.
}
\email{tod@maths.ox.ac.uk}
\maketitle

\maketitle
\begin{abstract}
We construct isometric and conformally isometric embeddings  of some gravitational instantons
in $\R^7$ and  $\R^8$. In particular we show that the embedding  class of the Einstein--Maxwell
instanton due to Burns is equal to $3$. For $\CP^2$, Eguchi--Hanson and anti-self-dual Taub-NUT
we obtain  upper and lower bounds on the embedding class.
\end{abstract}
\section{Introduction}
Gravitational instantons are solutions to the four-dimensional Einstein equations
in Riemannian signature which give complete metrics and are compact, or asymptotically
look like flat space. Some instantons (for example the Euclidean Schwarzchild metric) arise as analytic continuations
of Lorentzian metrics, but those with anti-self-dual conformal curvature do not have Lorentzian 
analogues \cite{Gibbons:1979zt, Dbook}.
In an attempt to visualise gravitational instantons one may isometrically embed them in a flat higher--dimensional
ambient space as surfaces.

 In this paper we shall consider such embeddings as well as conformal embeddings (see
Definition \ref{defi1}) where an instanton embeds in a conformally flat space. We shall exhibit explicit embeddings
of Eguchi--Hanson, anti--self--dual Taub NUT, and the Fubini--Study metrics. In all three cases
the embedding class (see Definition \ref{deficlass}) is at most $4$, and at least $3$, and the conformal embedding class is at most $3$ and at least $2$. The only case (apart from the Schwarzschild instanton, where the embedding class is known to be $2$) where we could establish a sharp result
about the embedding class is an Einstein--Maxwell instanton known as the Burns metric \cite{burns_ref}. 
It is a scalar--flat K\"ahler metric  on the total space of the line bundle ${\mathcal O}(-1)\rightarrow\CP^1$ (see (\ref{burns_met}) for the coordinate formula)
\begin{theo}
\label{theo_burns}
The isometric embedding class of the Burns metric is $3$.
\end{theo}
The paper is organised as follows. In the next section we shall recall the basic theory of isometric embeddings of class 2, and follow 
\cite{russian, agaoka} to give (in Theorem \ref{theoAY}) necessary conditions for an
existence of class 2 embeddings of four--manifolds. 
In \S\ref{section_burns} we shall introduce the Burns metric, and establish Theorem \ref{theo_burns}.
In \S\ref{sectioncp2} we shall use the conformal equivalence between the Fubini--Study metric and the Burns metric to rule out the existence of a radially
symmetric
 conformal embedding of $\CP^2$ in $\R^6$, and
show (Theorem \ref{theo52}) that the conformal embedding of $\CP^2$ in $\R^7$ via Burns is induced by
the canonical Veronese embedding $\CP^2\rightarrow S^7$. In \S\ref{sectionLRS} we shall construct explicit
conformal embeddings of the Eguchi--Hanson and anti--self--dual (ASD) Taub-NUT gravitational instantons
in $\R^7$. The resulting embedding is global in the case of ASD Taub-NUT. We shall prove
(Theorem \ref{theounib}) that the Burns metric is  the unique non--flat locally
rotationally symmetric Bianchi IX metric which is ASD and scalar--flat and which admits
a radial isometric embedding in $\R^7$. In \ref{sub71} we shall construct isometric embeddings of
Eguchi--Hanson and ASD Taub--NUT in $\R^8$.
\subsection*{Acknowledgements.} We are grateful 
to Robert Bryant for correspondence, and pointing out the reference \cite{agaoka} to us. The work of M.D. has been partially supported by STFC consolidated grant no. ST/P000681/1.

\section{Isometric embeddings of class 2}
Let $\R^{r, s}$ be an $(r+s)$--dimensional pseudo--Euclidean space with a flat metric $\eta$
of signature $(r, s)$. We shall start off with a definition 
\begin{defi}
\label{defi1}
An isometric embedding of a pseudo--Riemannian $n$--dimensional manifold
$(M, g)$ as a surface in $\R^{r, s}$ is a map $\iota:M\rightarrow\R^{r, s}$ such that
$\iota^*(\eta)=g$ and
$\iota(M)\subset \R^{r, s}$ is diffeomorphic to $M$. 
\end{defi}
\noindent
In the real
analytic category an $n$--dimensional manifold is always locally embeddable in $\R^N$,
where $N=n(n+1)/2$ dimensions \cite{cartan}. 
\begin{defi}
\label{deficlass}
The isometric embedding class of $(M, g)$ is the smallest integer $k$ such that
there exists an isometric embedding of $(M, g)$ into $\R^{r, s}$ with $r+s-n=k$.
\end{defi}
\noindent
An  example of a gravitational instanton with embedding class $2$
is the Euclidean Schwarzchild metric
\[
g=\Big(1-\frac{2m}{r}\Big)d\tau^2+\Big(1-\frac{2m}{r}\Big)^{-1}dr^2+r^2(d\theta^2+\sin{\theta}^2 d\phi^2) \quad
r>2m.
\]
If  $\tau$ is periodic with the period $8\pi m$, then $g$
can be globally and isometrically embedded in $\R^6$ with the flat metric
\[
\eta=d\rho^2+\rho^2d\psi^2+dZ^2+dR^2+R^2 (d\Theta^2+\sin{\Theta}^2 d\Phi^2).
\]
The embedding is given by a modification of the Fronsdal construction \cite{fronsdal}
\be
\label{schw_emb}
 R=r,\quad \rho=4m\sqrt{1-\frac{2m}{r}}, \quad 
Z=\int\sqrt{\Big(\frac{2m}{r}\Big)^3 + \Big(\frac{2m}{r}\Big)^2+\Big(\frac{2m}{r}\Big)}dr,\quad \psi=\frac{\tau}{4m}, 
\quad \Theta=\theta, \quad \Phi=\phi.
\ee
The corresponding surface (Figure \ref{dt})
in $\R^6$ is approximated by a paraboloid $8mZ=\sqrt{3}\rho^2 $ 
near $r=2m$, and for large $r$ it asymptotically approaches the flat
cylinder $\rho=4m$. 
\begin{center}
\label{dt}
\includegraphics[width=5cm,height=5cm,angle=0]{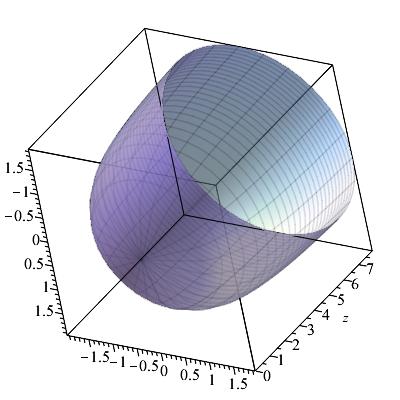}
\begin{center}
{\em {\bf Figure 1.} The surface $M$ (\ref{schw_emb}) in $(\rho, \psi, Z)$ coordinates.  
Each point on $M$ corresponds to a two--sphere.}
\end{center}
\end{center}
Moreover, as $g$ is Ricci--flat, there does not exist an embedding in $\R^5$,
even locally \cite{exsol}. The embedding class of the Euclidean Schwarschild metric is therefore $2$.

In the rest of this Section we shall review some necessary conditions
for existence of class 2 isometric embeddings of an oriented Riemannian
four--manifold $(M, g)$. Let $\epsilon_{abcd}$ and ${R_{abc}}^d$ be the
volume form and the Riemann tensor of $g$ respectively. Set
\be\label{a1}P_{ab}^{\;\;\;\;cd}:=-\frac12\epsilon_{abpq}R^{pqcd},\ee
and define
\be
\label{a2}
\Phi_3\equiv \mbox{Tr}(P^3)= {P^{ab}}_{cd} {P^{cd}}_{ef} {P^{ef}}_{ab}, \quad
\Phi_5\equiv \mbox{Tr}(P^5).
\ee
\begin{theo}[Yakupov \cite{russian}, Agaoka \cite{agaoka}]
\label{theoAY}
Necessary conditions for an existence of a local isometric
embedding of a four--manifold $(M, g)$ into $\R^{r, s}$ with $r+s=6$ are
\be
\label{aga_conds}
\Phi_3=0, \quad \Phi_5=0.
\ee
\end{theo}
The sketch of proof given in \cite{russian} applies only to Einstein manifolds, while Agaoka's proof \cite{agaoka} is relatively
long. In \S \ref{section2} we shall give a self--contained proof of this Theorem 
based on the Gauss equation.
The conditions (\ref{aga_conds}) are not sufficient for local class
2 embeddability, and further local obstructions arise. Some of these
are of higher differential order, and result from the Codazzi equations \cite{kaneda} but
in \S\ref{section21} we shall show that the Gauss equation alone may not be satisfiable
even if (\ref{aga_conds}) holds, as it may not be possible to construct
real second--fundamental forms of the embeddings.
\subsection{Basic Theory}
\label{section2}
Let $\iota:M\rightarrow \R^{r, s}$ be a class two isometric embedding, with
$r\geq 4$ and $r+s=6$.  The embedded surface $M$ has two normals so two second fundamental forms but also a vector defining the connection on the normal bundle. Basic theory can be found for example in \cite{exsol}, but here is another way. We shall first
deal with Riemannian $\R^6$, and then incorporate other signatures. Choose an orthonormal basis in 6 dimensions containing the 2 normals along with 4 vectors tangent to $M$ and write the 6dim coordinate vector
as
\[X^\alpha=(X^a,Y,Z),\]
so $\alpha=0,\ldots,5$ while $a=0,\ldots,3$, and
\[\delta_{\alpha\beta}X^\alpha X^\beta=g_{ab}X^aX^b+Y^2+Z^2,\]
with $\delta_{\alpha\beta}$ the flat metric of $\mathbb{R}^6$ and $g_{ab}$ the induced metric of $M$.
\begin{prop}
There exists symmetric tensors $K_{ab}, L_{ab}$, and a one--form $V_a$ such that
\be\label{e7}R_{abcd}=-2K_{c[a}K_{b]d}-2L_{c[a}L_{b]d},
\quad\mbox{Gauss equation}
\ee
\be\label{e8}\nabla_{[a}K_{b]c}=V_{[a}L_{b]c}, \quad
\nabla_{[a}L_{b]c}=-V_{[a}K_{b]c},
\quad\mbox{Codazzi equations}
\ee
\be
\label{e11}\nabla_{[a}V_{b]}=L_{c[a}K_{b]}^{\;\;\;c}
\quad
\mbox{Ricci equation.}
\ee
\end{prop}
\noindent
{\bf Proof.}
The symmetric tensors (the \emph{second fundamental forms}) $K_{ab},L_{ab}$ 
on $M$ arise from
\be\label{e1}\nabla_aX_b=g_{ab}-YK_{ab}-ZL_{ab},\ee
and a vector (the \emph{torsion vector}) $V_a$ satisfies
\be\label{e4}\nabla_aY=K_{ab}X^b+ZV_a,\ee
\be\label{e5}\nabla_aZ=L_{ab}X^b-YV_a.\ee
The formalism invites a complexification which emphasises the role of $V_a$ in defining a connection on the normal bundle: set
\[\zeta=Y+iZ,\;\;\Sigma_{ab}=K_{ab}+iL_{ab},\]
when
\be\label{e6}\nabla_a\zeta=\Sigma_{ab}X^b-i\zeta V_a.\ee
Now a rotation in the normal bundle has the effect
\[\zeta\rightarrow\hat{\zeta}=e^{i\Theta(X)}\zeta,\;\;\Sigma_{ab}\rightarrow\hat{\Sigma}_{ab}=e^{i\Theta}\Sigma_{ab},\;\;V_a\rightarrow\hat{V}_a=e^{i\Theta}(V_a-\nabla_a\Theta).\]

\medskip

\noindent Commute derivatives on $X_a,Y,Z$ to obtain (\ref{e7}, \ref{e8}, \ref{e10}), and hence also

\be\label{e10}\nabla_{[a}\Sigma_{b]c}=-iV_{[a}\Sigma_{b]c}.\ee
Note that the consistency conditions for (\ref{e8}) are satisfied by virtue of themselves and (\ref{e11}), and that for (\ref{e11}) is automatically satisfied by virtue of (\ref{e8}). The Bianchi identity from (\ref{e7}) is automatic given (\ref{e8}).
\koniec
If we want embedding into signatures other than Riemannian we may replace the above by
\[\delta_{\alpha\beta}X^\alpha X^\beta=g_{ab}X^aX^b+\epsilon_1Y^2+\epsilon_2Z^2,\]
with $\epsilon_i=\pm1$. Then the second fundamental forms occur in the modified expression
\be\label{e22}\nabla_aX_b=g_{ab}-\epsilon_1YK_{ab}-\epsilon_2ZL_{ab},\ee
and the vector $V_a$ in
\[\nabla_aY=K_{ab}X^b+ZV_a,\]
\[\nabla_aZ=L_{ab}X^b-\epsilon_1\epsilon_2YV_a.\]
The Gauss equation becomes
\be\label{e17}R_{abcd}=-2\epsilon_1K_{c[a}K_{b]d}-2\epsilon_2L_{c[a}L_{b]d},\ee
the Codazzi equations can be taken to be
\be\label{e18}\nabla_{[a}K_{b]c}=V_{[a}L_{b]c}, \quad
\nabla_{[a}L_{b]c}=-\epsilon_1\epsilon_2V_{[a}K_{b]c},\ee
and the Ricci equation becomes
\be\label{e21}\nabla_{[a}V_{b]}=\epsilon_2L_{c[a}K_{b]}^{\;\;\;c}.\ee
\subsection{Necessary conditions}
We shall mostly be concerned with solving the Gauss equation (\ref{e7}) in what follows, and it will be sufficient to establish Theorem 
\ref{theoAY}\\
{\bf Proof of Theorem \ref{theoAY}}. First note an identity for any symmetric $T^{ab}$ in 4 dimensions:
\be\label{a3}\epsilon_{abpq}T^{pm}T^{qn}\epsilon_{mnef}T^{er}T^{fs}\epsilon_{rscd}=4(\mbox{det}T)\epsilon_{abcd},\ee
(this is in Riemannian signature; there will be sign modifications in Lorentzian signature). The determinant is defined by contraction of this formula with $\epsilon^{abcd}$. The identity is easy to see for diagonal $T^{ab}$, and in Riemannian signature all symmetric $T$ can be diagonalised.

Introduce a matrix notation: for symmetric $T^{ab}$ the tensor $X_{ab}^{\;\;\;\;cd}:=\epsilon_{abpq}T^{pc}T^{qd}$ defines an endomorphism on 2-forms, which is necessarily trace-free.  Since
\[X_I^{\;\;J}=X_{ab}^{\;\;\;cd}=\epsilon_{abef}T^{ec}T^{fd},\]
we need 
\be\label{ec1}X_{ab}^{\;\;\;cb}=0\ee 
by the symmetry of $T^{ab}$.
 Suppress the indices and use bold font to indicate a $6\times 6$-matrix, then (\ref{a3}) can be written
\be\label{a4}{\bf X}^2=4(\mbox{det}T){\bf I},\ee
where, as a tensor, ${\bf I}=\delta_{[a}^{\;\;c}\delta_{b]}^{\;\;d}$, and recall that $\mbox{tr}{\bf X}=0$ by the symmetry of $T^{ab}$, so that (\ref{a4}) is in fact the minimum polynomial of ${\bf X}$ (assuming $\bX$ is nontrivial).

The Gauss equation (\ref{e7}) shows that, as a matrix,
\[{\bf P}={\bf X}+{\bf Y},\]
where ${\bf X},{\bf Y}$ are made from $K^{ab},L^{ab}$ respectively in the place of $T^{ab}$.

To obtain (\ref{aga_conds}) we need to calculate
\[\Phi_3=\mbox{tr}(({\bf X}+{\bf Y})^3).\]
Note from (\ref{a4}) that
\[\mbox{tr}({\bf X}^3)=4(\mbox{det}K)\mbox{tr}({\bf X})=0,\]
and
\[\mbox{tr}({\bX}^2\bY)=4(\mbox{det}K)\mbox{tr}({\bf Y})=0,\]
whence $\Phi_3=0$.
The second condition $\Phi_5=0$  is the vanishing of $\mbox{tr}(({\bf X}+{\bf Y})^5)$ and for this we need to consider
\[\mbox{tr}({\bf X}^5)=16(\mbox{det}K)^2\mbox{tr}({\bf X})=0,\]
\[\mbox{tr}({\bf X}^4{\bf Y})=16(\mbox{det}K)^2\mbox{tr}({\bf Y})=0,\]
\[\mbox{tr}({\bf X}^3{\bf Y}^2)=16(\mbox{det}K\mbox{det}L)\mbox{tr}({\bf X})=0,\]
\[\mbox{tr}({\bf X}^2{\bf Y}{\bf X}{\bf Y})=16(\mbox{det}K)\mbox{tr}({\bf Y}{\bf X}{\bf Y})=0,\]
from which the result follows.

The Cayley-Hamilton Theorem for $\bP$ shows the vanishing of $\mbox{tr}(\bP^{2k+1})$ inductively for all odd powers, since the characteristic polynomial of ${\bf P}$ must take the form
\[{\bf P}^6+c_1{\bf P}^4+c_2{\bf P}^2+c_3{\bf I}=0.\]
\koniec
We can now rule out the existence of class 2 isometric embeddings
for all anti--self--dual gravitational instantons, as well as for the
Kerr solution
\begin{col}
\label{col1}
None of $\mathbb{CP}^2$ with the Fubini-Study metric, the (Lorentzian or Riemannian) Kerr solution, or any  hyper--K\"ahler metric  
in four--dimensions with the $J$--invariant not equal to zero
have isometric embedding class 2.
\end{col}\noindent
{\bf Proof.}
Recall \cite{pr} the spinor decomposition of the Weyl tensor
\[
C_{abcd}=\psi_{ABCD}\epsilon_{A'B'}\epsilon_{C'D'}+\tilde{\psi}_{A'B'C'D'}\epsilon_{AB}\epsilon_{CD},
\]
where $\psi, \tilde{\psi}$
are totally symmetric
in their indices, and the $\epsilon$ spinors are skew.
There are four algebraic invariants of the Weyl spinors:
\begin{eqnarray}
\label{IJ}
I&=&\psi_{ABCD}\psi^{ABCD}, \quad J={\psi_{AB}}^{CD}{\psi_{CD}}^{EF}{\psi_{EF}}^{AB},\\
\tilde{I}&=&\psi_{A'B'C'D'}\psi^{A'B'C'D'}, \quad \tilde{J}={\psi_{A'B'}}^{C'D'}{\psi_{C'D'}}^{E'F'}{\psi_{E'F'}}^{A'B'},\nonumber
\end{eqnarray}
which are in general independent. For an Einstein metric
\[
P_{ab}^{\;\;\;\;cd}=-\frac{1}{2}\epsilon_{abef}C^{efcd}-\frac{s}{12}\epsilon_{ab}^{\;\;\;\;cd},
\]
where $s$ is the scalar curvature. Therefore
\be\label{a4a} \Phi_3=8(J-\tilde{J})+2s(I-\tilde{I}).
\ee
We can now rule out several possibilities: The Fubini--Study metric
on $\CP^2$ is conformaly ASD (see \S\ref{sectioncp2}), and we find $\Phi_3=-s^3/9$. This
was established in \cite{agaoka}. Any hyper--K\"ahler metric in four--
dimensions is ASD and Ricci--flat. Therefore $\tilde{I}=\tilde{J}=s=0$.
Therefore $\Phi_3\neq 0$ iff $J\neq 0$. For the Lorentzian 
Kerr solutions $J$ is complex and 
Lorentzian signature $\tilde{J}=\overline{J}$, so $\Phi_3\neq 0$. In Riemannian 
signature 
$J\sim (r-\alpha\cos{\theta})^{-9}$ and $\widetilde{J}\sim (r+\alpha\cos{\theta})^{-9}$
in the natural Kerr coordinates \cite{EGH} , therefore $J-\widetilde{J}\neq 0$.
\koniec
\subsection{Failure of sufficiency}
\label{section21}
Assume that the Riemann curvature of $(M, g)$ satisfies the conditions
(\ref{aga_conds}) of Theorem (\ref{theoAY}). In order to construct the embedding we should be able to 
\begin{enumerate}
\item Find ${\bf X}$ and ${\bf Y}$ such that
\be
\label{XYfromP}
{\bf P}={\bf X}+{\bf Y},\;\;{\bf X}^2=\lambda{\bf I},\;\;{\bf Y}^2=\mu{\bf I},\;\;\mbox{tr}{(\bf X)}=0= \mbox{tr}{(\bf Y)},
\ee
where $\lambda, \mu$ are also to be found.
\item Assuming that ${\bf X}$ and ${\bf Y}$
satisfying (\ref{XYfromP}) 
 have been found, find symmetric tensors
$K$ and $L$ such that
\be\label{k1}X_{ab}^{\;\;\;\;cd}:=\epsilon_{abpq}K^{pc}K^{qd},\quad
Y_{ab}^{\;\;\;\;cd}:=\epsilon_{abpq}L^{pc}L^{qd}.
\ee
\item Assuming that $K$ and $L$ satisfying (\ref{k1})
have been found, check that they satisfy
the Codazzi equations (\ref{e8}).
\end{enumerate}
In fact each of these steps may fail. We shall first show that
the second step can fail in general, that is  real second fundamental forms
need not arise from real ${\bf X}$ and ${\bf Y}$. In \S\ref{section_burns} 
we shall then use the
example of the Burns metric to show that the first step can also fail.
\subsection{Obtaining $K_{ab}$ from $\bX$ in the general case}
\begin{prop}
Conditions (\ref{aga_conds})  are not sufficient for the existence of  symmetric tensors $K$ and  $L$
such that the Gauss equation (\ref{e7}) holds.
\end{prop}
\noindent
{\bf Proof.}
If we had found real $\bX,\bY$, we would still need to find real $K_{ab},L_{ab}$ and there are new complex issues here. Suppose then we have a candidate real $\bX$ with
\[\mbox{tr}({\bf X})=0,\;\;\bX^2=4\sigma_4\bI,\;\;X_{acb}^{\;\;\;\;\;c}=0,\]
for real constant $\sigma_4$, which will be $\mbox{det}K$,  can we find real symmetric $K_{ab}$ satisfying (\ref{k1})?
Introduce
\[T_{ab}=\epsilon_{adef}X_b^{\;\;def}.\] Then $T_{ab}$ is symmetric by virtue of the conditions on $\bX$ and (\ref{k1}) becomes
\be\label{k2}T_{ab}=K_a^{\;c}K_{bc}-\sigma_1K_{ab},\ee
with $\sigma_1=K_c^{\;\;c}$. Again introduce a matrix notation, but now for $4\times 4$-matrices:
\[\bK:=(K_{ab}),\;\;\bT=(T_{ab}),\]
and let $\sigma_i$ for $i=1,\ldots,4$ be the symmetric polynomials of $\bK$, so that
\[\sigma_1=\mbox{tr}\bK, \sigma_4=\mbox{det}\bK,\]
\[\mbox{tr}(\bK^2)=\sigma_1^2-2\sigma_2,\]
\[\mbox{tr}(\bK^3)=\sigma_1^3-3\sigma_1\sigma_2+3\sigma_3,\]
\[\mbox{tr}(\bK^4)=\sigma_1^4-4\sigma_1^2\sigma_2+2\sigma_2^2+4\sigma_1\sigma_3-4\sigma_4,\]
and the Cayley-Hamilton Theorem for $\bK$ is
\[\bK^4-\sigma_1\bK^3+\sigma_2\bK^2-\sigma_3\bK+\sigma_4\bI=0.\]
Given the Cayley-Hamilton Theorem we may calculate
\[\mbox{tr}(\bK^5)= \sigma_1^5-5\sigma_1^3\sigma_2 +5\sigma_1^2\sigma_3+5\sigma_1\sigma_2^2-5\sigma_2\sigma_3-5\sigma_1\sigma_4,\]
and
\[\mbox{tr}(\bK^6)= \sigma_1^6-6\sigma_1^4\sigma_2 +6\sigma_1^3\sigma_3+9\sigma_1^2\sigma_2^2-6\sigma_1^2\sigma_4 -12\sigma_1\sigma_2\sigma_3+3\sigma_3^2+6\sigma_2\sigma_4. \]
In matrix notation, (\ref{k2}) becomes
\be\label{k3}\bT=\bK^2-\sigma_1\bK,\ee
and we wish to solve this for real symmetric $\bK$ given real symmetric $\bT$. Since $\bK$ is real symmetric, there will be real orthogonal $\bO$ and real diagonal $\bDelta$ with
\[\bK=\bO^T\bDelta\bO,\]so (\ref{k3}) becomes
\[\bT=\bO(\bDelta^2-\sigma_1\bDelta)\bO^T,\]
and therefore the same $\bO$ diagonalises $\bT$. Given $\bT$ we find $\bO$, then we just need the $\sigma_i$ to fix $\bDelta$. We already have
\[\sigma_4=\frac{1}{24}\mbox{tr}(\bX^2).\]
From (\ref{k3}) by taking the trace we obtain
\[\sigma_2=-\frac12\mbox{tr}(\bT).\]
By squaring and tracing (\ref{k3}) we find
\be\label{k4}\sigma_1\sigma_3 =\sigma_2^2-2\sigma_4-\frac12\mbox{tr}(\bT^2),\ee
and by cubing and tracing (\ref{k3})
\be\label{k5}\mbox{tr}(\bT^3)=5\sigma_1^2\sigma_4+3\sigma_1\sigma_2\sigma_3+3\sigma_3^2+6\sigma_2\sigma_4.\ee
Eliminate $\sigma_1$ in favour of $\sigma_3$ with the aid of (\ref{k4}), then (\ref{k5}) gives a quadratic for $\sigma_3^2$ (What if the expression on the right in (\ref{k4}) vanishes? Then one of $\sigma_1,\sigma_3$ vanishes; one chooses which and then (\ref{k5}) gives the other, up to sign -- the solution is non-unique but possibly complex). There is a sign ambiguity in $\sigma_3$ and therefore also in $\sigma_1$ but this is to be expected as $K_{ab}$ can only be known up to sign, but, starting from $\bX$, there is no reason for $\sigma_3$ or $\sigma_1$ to be real, so reality of these is an extra condition on $\bT$ and therefore on $\bX$. To see that this is a genuine issue, note that $\bK=i\bI$ leads to a real $\bT$, so indeed a real $\bT$ is not obliged to give a real $\bK$.
\koniec
\section{The Burns metric}
\label{section_burns}
In this Section  we shall introduce the Burns metric \cite{burns_ref}.
It is K\"ahler with vanishing scalar curvature (and therefore has
anti--self--dual Weyl tensor), and it is one of the few metrics
where the embedding class can be established as in Theorem \ref{theo_burns}.

The Burns metric is the unique
scalar--flat K\"ahler metric  on the total space of the line bundle 
$\mathcal{O}(-1)\rightarrow \CP^1$. It can also be regarded 
as an Einstein--Maxwell gravitational instanton, with the self--dual
part of the Maxwell field strength given by the K\"ahler form \cite{DH}. 

There are several ways to present it in 
coordinates, and we chose to write it as 
\be
\label{burns_met}
g_B=d\rho^2+\frac{1}{4}\rho^2 (\sigma_1^2+\sigma_2^2+\sigma_3^2)+\frac{1}{4}(\sigma_1^2+\sigma_2^2),
\ee
where $\sigma_1, \sigma_2, \sigma_3$ are
left--invariant one
forms on $SU(2)$, such that $d\sigma_1+\sigma_2\wedge\sigma_3=0$ etc.
We shall verify by explicit computation that the conditions
\ref{aga_conds} of Theorem \ref{theoAY} are satisfied for this metric, and yet
there does not exist a class 2 isometric embedding as real
{\bf X} does not exist. We shall also establish
a (rather obvious) embedding in $\R^7$. This will complete the proof of 
Theorem \ref{theo_burns}.\\
{\bf Proof of Theorem \ref{theo_burns}}
We will first show that the (\ref{aga_conds}) hold for the Burns metric, and 
then demonstrate that these conditions are not sufficient for the existence
of class 2 embedding. The vanishing of $\Phi_3$ and $\Phi_5$ could be
verified by a MAPLE computation given the form of the metric (\ref{burns_met})
but we find it instructive to present a computation which can be directly 
checked by readers, and exhibits the expicit form of ${\bf P}$.

To calculate the curvature, use Cartan calculus with the orthonormal tetrad:
\be\label{bm2}\theta^0=d\rho,\;\;\theta^1=\frac12(1+\rho^2)^{1/2}\sigma^1,\;\;\theta^2=\frac12(1+\rho^2)^{1/2}\sigma^2,\;\;\theta^3=\frac{\rho}{2}\sigma^3.\ee
The connection 1-forms are:
\[\omega^0_{\;\;1}=-\frac{\rho}{1+\rho^2}\theta^1,\;\;\omega^0_{\;\;2}=-\frac{\rho}{1+\rho^2}\theta^2,\;\;\omega^0_{\;\;3}=-\frac{1}{\rho}\theta^3,\]
\[\omega^1_{\;\;2}=\frac{\rho^2+2}{\rho(1+\rho^2)}\theta^3,\;\;\omega^3_{\;\;1}=\frac{\rho}{(1+\rho^2)}\theta^2,\;\;\omega^2_{\;\;3}=\frac{\rho}{(1+\rho^2)}\theta^1,\]
leading to Riemann tensor components
\[\label{bm4}R_{0101}=R_{0202}=-R_{0123}=-R_{0231}=R_{3131}=R_{2323}=-(1+\rho^2)^{-2},\]
\[\label{bm5}R_{0312}=-2(1+\rho^2)^{-2},\;\;R_{1212}=4(1+\rho^2)^{-2}.
\]
The Ricci components are
\[R_{00}=-R_{11}=-R_{22}=R_{33}=-2(1+\rho^2)^{-2},\]
and the scalar curvature is zero (as expected).

Now consider $P_I^{\;\;J}\equiv -P_{ab}^{\;\;\;\;\;\;cd}=\frac{1}{2}\epsilon_{abpq}R^{pqcd}$ as a $6\times 6$ matrix with the indices ordered
\[I,J=01,23,02,31,03,12.\]
Claim this is
\be\label{bm6}\bP= \left(\begin{array}{rrrrrr}
           -1 & 1 & 0 & 0 & 0 & 0\\
          1 & -1 & 0 & 0 & 0 & 0\\
           0 & 0 & -1 & 1 & 0 & 0\\
           0 & 0 & 1 & -1 & 0 & 0\\
           0 & 0 & 0 & 0 & 2 & -4\\
           0 & 0 & 0 & 0 & 0 & 2\\
\end{array}\right), \ee
omitting the factor $(1+\rho^2)^{-2}$ which is common to all terms. This can be written
\[\bP=\mbox{diag}\left(\left(\begin{array}{rr}
                      -1&1\\
                      1&-1\\
                     \end{array}\right),\left(\begin{array}{rr}
                      -1&1\\
                      1&-1\\
                     \end{array}\right),\left(\begin{array}{rr}
                      2&-4\\
                      0&2\\
                     \end{array}\right)\right),\]

and this form is easier for calculation. One finds

\[\bP^3=\mbox{diag}\left(\left(\begin{array}{rr}
                      -4&4\\
                      4&-4\\
                     \end{array}\right),\left(\begin{array}{rr}
                      -4&4\\
                      4&-4\\
                     \end{array}\right),\left(\begin{array}{rr}
                      8 & -48\\
                      0& 8\\
                     \end{array}\right)\right),\]

\[\bP^5=\mbox{diag}\left(\left(\begin{array}{rr}
                      -16&16\\
                      16&-16\\
                     \end{array}\right),\left(\begin{array}{rr}
                      -16&16\\
                      16&-16\\
                     \end{array}\right),\left(\begin{array}{rr}
                      32 & -320\\
                      0& 32\\
                     \end{array}\right)\right),\]

Now it is clear that Agaoka's necessary conditions $\mbox{tr}\bP^3=0=\mbox{tr}\bP^5$ both hold.
\vskip5pt

If the conditions (\ref{aga_conds}) of Theorem \ref{theoAY}  
were sufficient we should now be able to find the isometric embedding of the Burns metric and so in particular to solve the Gauss equation. This requires finding $\bX, \bY$  satisfying (\ref{XYfromP}). However, there is another condition on $\bX,\bY$, as (\ref{ec1}) holds automatically for $\bP=\bX+\bY$ but would not hold necessarily for the summands $\bX,\bY$ separately unless we imposed it.

We may seek $\bX$ in partitioned form as
\[\bX=\left(\begin{array}{ccc}
                       A&B&C\\
                      D&E&F\\
                      J&K&L\\
                    \end{array}\right)\]
where all blocks are $2\times 2$, and decompose $\bP$ for brevity as
\[\bP=\left(\begin{array}{ccc}
                       M&0&0\\
                      0&M&0\\
                      0&0&N\\
                    \end{array}\right),\]
with
\[M= \left(\begin{array}{rr}
                      -1&1\\
                      1&-1\\
                     \end{array}\right) ,\;\;N= \left(\begin{array}{rr}
                      2&-4\\
                      0&2\\
                     \end{array}\right) .\]
Write $\bY=\bP-\bX$ and square
\[\bY^2=\mu \bI=\bP^2-(\bP\bX+\bX\bP)+\lambda \bI,\]
or
\[
\bP\bX+\bX\bP=\bP^2+(\lambda-\mu)\bI={\bP}^2+k{\bI}\mbox{  say},
\]
with $k=\lambda-\mu$.
Now look at this in blocks:
\begin{eqnarray*}
\bP\bX+\bX\bP
                    &=&\left(\begin{array}{ccc}
                       AM+MA&BM+MB&CN+MC\\
                      DM+MD&EM+ME&FN+MF\\
                      JM+NJ&KM+NK&LN+NL\\
                    \end{array}\right)\\
                    &=&{\bP}^2+k{\bI}\\
&=&                    \left(\begin{array}{ccc}
                       M^2+kI&0&0\\
                      0&M^2+kI&0\\
                      0&0&N^2+kI\\
                    \end{array}\right).
										\end{eqnarray*}
This leads to constraints on the blocks in $\bX$ which we can solve successively. First solve $BM+MB=0$: set $B=\left(\begin{array}{cc}
                       \alpha&\beta\\
                      \gamma&\delta\\
                    \end{array}\right)$ then
\[MB+BM=\left(\begin{array}{cc}
                       -2\alpha+\beta+\gamma&-2\beta+\alpha+\delta\\
                      -2\gamma+\alpha+\delta&-2\delta+\beta+\gamma\\
                    \end{array}\right)=0.\]
This rapidly leads to
\[B=\left(\begin{array}{cc}
                       \beta&\beta\\
                      \beta&\beta\\
                    \end{array}\right),\]
                    for some $\beta$, and similarly
\[D=\left(\begin{array}{cc}
                       \delta&\delta\\
                      \delta&\delta\\
                    \end{array}\right),\]
                    for some $\delta$.
Next solve $CN+MC=0$: set $C=\left(\begin{array}{cc}
                       \alpha&\rho\\
                      \gamma&\sigma\\
                    \end{array}\right)$ then
\[CN+MC=\left(\begin{array}{cc}
                       \alpha+\gamma&-4\alpha+\rho+\sigma\\
                      \alpha+\gamma&-4\gamma+\rho+\sigma\\
                    \end{array}\right)=0,\]
                    whence
\[C=\left(\begin{array}{cc}
                       0&\gamma\\
                      0&-\gamma\\
                    \end{array}\right),\]
                    for some $\gamma$, and similarly
\[F=\left(\begin{array}{cc}
                       0&\phi\\
                      0&-\phi\\
                    \end{array}\right),\]
                    for some $\phi$. 
Next solve $JM+NJ=0$: set $J=\left(\begin{array}{cc}
                       \rho&\sigma\\
                      \tau&\nu\\
                    \end{array}\right)$ then
\[JM+NJ=\left(\begin{array}{cc}
                       -4\tau+\rho+\sigma&-4\nu+\rho+\sigma\\
                      \tau+\nu&\tau+\nu\\
                    \end{array}\right)=0,\]
                    whence
\[J=\left(\begin{array}{cc}
                       \rho&-\rho\\
                      0&0\\
                    \end{array}\right),\]
                    for some $\rho$ and 
\[K=\left(\begin{array}{cc}
                       \kappa&-\kappa\\
                      0&0\\
                    \end{array}\right),\]
                    for some $\kappa$.
For $A,E,L$ we have inhomogeneous equations (and we will re-use Greek letters). With $A=\left(\begin{array}{cc}
                       \alpha&\beta\\
                      \gamma&\delta\\
                    \end{array}\right)$ we want
\[AM+MA=\left(\begin{array}{cc}
                       -2\alpha+\beta+\gamma&-2\beta+\alpha+\delta\\
                      -2\gamma+\alpha+\delta&-2\delta+\beta+\gamma\\
                    \end{array}\right)=\left(\begin{array}{cc}
                       2+k&-2\\
                      -2&2+k\\
                    \end{array}\right),\]
                    which only has solutions with $k=0$ i.e. $\lambda=\mu$, and then
\[A=\left(\begin{array}{cc}
                       \alpha&\alpha+1\\
                      \alpha+1&\alpha\\
                    \end{array}\right),\]
                    for some $\alpha$, when also
\[E=\left(\begin{array}{cc}
                       \epsilon&\epsilon+1\\
                      \epsilon+1&\epsilon\\
                    \end{array}\right),\]
                    for some $\epsilon$.
Finally we need $L=\left(\begin{array}{cc}
                       \alpha&\beta\\
                      \gamma&\delta\\
                    \end{array}\right)$ satisfying
\[LN+NL=\left(\begin{array}{cc}
                       -4\gamma+4\alpha&-4\alpha-4\delta+4\beta\\
                      4\gamma&-4\gamma+4\delta\\
                    \end{array}\right)=\left(\begin{array}{cc}
                       4&-16\\
                      0&4\\
                    \end{array}\right)\]
                    since we now know that $k=0$, and this leads to
\[L=\left(\begin{array}{cc}
                       1&-2\\
                      0&1\\
                    \end{array}\right).\]
We have found $\bX$ in terms of parameters $\alpha,\beta,\gamma,\delta,\epsilon,\kappa,\phi$ and $\rho$, explicitly
\[\bX=
\left(\begin{array}{cccccc}
           \alpha & \alpha+1 & \beta & \beta & 0 & \gamma\\
          \alpha+1 & \alpha & \beta & \beta & 0 & -\gamma\\
           \delta & \delta & \epsilon & \epsilon+1 & 0 & \phi\\
           \delta & \delta & \epsilon+1 & \epsilon & 0 & -\phi\\
           \kappa & -\kappa & \rho & -\rho & 1 & -2\\
           0 & 0 & 0 & 0 & 0 & 1\\
\end{array}\right). \]
For the vanishing trace we need $\alpha+\epsilon=-1$, which we impose, then we impose $\bX^2=\lambda \bI$ for some $\lambda$. By inspection of $\bX$ (consider the bottom right entry of $\bX^2$) only $\lambda=1$ is possible. For the rest we obtain the conditions
\[\alpha(\alpha+1)+\beta\delta=0=\epsilon(\epsilon+1)+\beta\delta,\]
and
\be\label{cc1}\kappa\gamma+\rho\phi=1.\ee
Given $\alpha+\epsilon+1=0$ from the vanishing trace, only one of the first pair is independent, so there are 6 free parameters remaining in $\bX$, and $\bY$ is uniquely determined given $\bX$. 

However we still need to impose (\ref{ec1}). Claim
\[X_{0b}^{\;\;\;1b}=\phi+\rho=0,\]
\[X_{0b}^{\;\;\;2b}=-\gamma-\kappa=0,\]
\[X_{0b}^{\;\;\;3b}=\beta-\delta=0,\]
and the rest give nothing new. Now we have $\phi=-\rho,\kappa=-\gamma$ which with (\ref{cc1}) gives
\[1=\kappa\gamma+\rho\phi=-\kappa^2-\phi^2,\]
which can't be satisfied with reals: there are no suitable real $\bX,\bY$. In particular the Burns metric is not isometric embedding class 2 into Riemannian $\mathbb{R}^6$ . By (\ref{e17}), embedding into other signatures still requires real $\bX$ and so these are also ruled out.
\vskip5pt
To complete the proof we should establish an isometric embedding of the Burns metric in $\R^7$, but this is readily done: the first
factor in (\ref{burns_met}) just flat $\R^4$, and the second factor is the round two--sphere embeddable in $\R^3$. To write this condition explicitly, as intersection of algebraic surfaces
consider a flat metric on $\R^7$
\[
\eta=|d\zeta_1|^2+|d\zeta_2|^2+{dx_1}^2+{dx_2}^2{+dx_3}^2,
\]
where
\[
x_1=Z\cos{\theta}\cos{\phi}, \quad x_2=Z\cos{\theta}\sin{\phi}, \quad
x_3=Z\sin{\theta}
\]
and
\[
\zeta_1= \rho\cos{\Big(\frac{\theta}{2}\Big)}e^{i(\psi+\phi)/2}, \quad
\zeta_2= \rho\sin{\Big(\frac{\theta}{2}\Big)}e^{i(\psi-\phi)/2}
\]
so one condition is
\be
\Big(\frac{|\zeta_1|^2-|\zeta_2|^2}{|\zeta_1|^2+|\zeta_2|^2}\Big)^2
=\frac{{x_3}^2}{{x_1}^2+{x_2}^2+{x_3}^2}.
\ee
We also note that ${\zeta_1}/{\zeta_2}=|{\zeta_1}/{\zeta_2}|e^{i\phi}$
so the second condition is
\[
i\frac{\zeta_1\bar{\zeta_2}+\bar{\zeta}_1\zeta_2}{ 
\zeta_1\bar{\zeta_2}-\bar{\zeta}_1\zeta_2}=\frac{x_1}{x_2}
\]
which looks like an orbifold in $\R^7$. The final condition is
\[
{{x_1}^2+{x_2}^2+{x_3}^2}=1/4.
\]
\koniec
\section{Conformal isometric embeddings 
of the Fubini--Study metric}
\label{sectioncp2}
Recall the definition \ref{defi1} of an isometric embedding, and
make the following definition \cite{jacobowitz, DT19}
\begin{defi}
A conformally isometric embedding of a pseudo--Riemannian $n$--dimensional manifold
$(M, g)$ as a surface in $\R^{r, s}$ is a map $\iota:M\rightarrow\R^{r, s}$ such that
$\iota^*(\eta)=\Omega^2 g$ for some  $\Omega: M\rightarrow \R^+$ and
$\iota(M)\subset \R^{r, s}$ is diffeomorphic to $M$. 
\end{defi}
\noindent
We shall also define the conformal embedding class of $(M,  g)$ to be the smallest integer $k$, such that $(M, g)$ can be conformally isometrically embedded in
$\R^{r, s}$ with $r+s-n=k$.

In this section we shall use the conformal equivalence
between the Burns metric and the Fubini--Study metric
on $\CP^2$ to show that the later does not admit a conformal embedding of 
class  2, under an additional assumption that the conformal factor is constant
on the $SU(2)$ orbits (the existence of the general conformal embedding of class
1 for $\CP^2$ was ruled out in \cite{DT19}).

The local form of the Fubini--Study metric is \cite{GP}
\be
\label{cp2metric}
g=\frac{dr^2}{(1+r^2)^2}+
\frac{1}{4}\frac{r^2\sigma_3^2}{(1+r^2)^2}
+\frac{1}{4}\frac{r^2}{1+r^2}(\sigma_1^2+\sigma_2^2).
\ee
The metric is regular everywhere on $\CP^2$, and 
the apparent singularity at $r=0$ results from 
using spherical polars.
Computing $\Phi_3$ given by (\ref{a2})
for (\ref{cp2metric}) gives  $-1536$ (the Ricci scalar is 
$24$ - if the Ricci scalar was $s$, then we would get $-s^3/9$). We are  {\em assuming} that a global conformal embedding (if one exists)
must have a conformal factor depending only of $r$. If we do not assume  this, then we are just restricting the class of conformal factors. 
Set $\hat{g} =\Omega^2 g$, where (for convenience) we set $\Omega=\exp{G(r)}$. Now we compute $\Phi_3$, and find that it vanishes iff
\be
\label{ODE}
-r \left( -3+ \left( {r}^{3}+r \right) {\frac {\rm d}{{\rm d}r}}G
 \left( r \right)  \right)  \left( {r}^{2}+1 \right) ^{2}{\frac {
{\rm d}^{2}}{{\rm d}{r}^{2}}}G \left( r \right) +{r}^{2} \left( {r}^{2
}+1 \right) ^{3} \left( {\frac {\rm d}{{\rm d}r}}G \left( r \right) 
 \right) ^{3}+
\ee
\[
 \left( -2\,{r}^{7}+6\,{r}^{3}+4\,r \right)  \left( {
\frac {\rm d}{{\rm d}r}}G \left( r \right)  \right) ^{2}+ \left( 3\,{r
}^{4}+12\,{r}^{2}+9 \right) {\frac {\rm d}{{\rm d}r}}G \left( r
 \right) -16\,r=0
\]
which is a first order ODE for $G'(r)$. 
We have shown that
$\Phi_5\equiv \mbox{Tr}(P^5)$ is another obstruction. Computing this for $\hat{g}$ gives another second order ODE
for $G(r)$ (this one also does not contain $G(r)$, but takes more space so we do not write it down).
We now solve (\ref{ODE}) for $G''$ and substitute to 
$\Phi_5=0$. This gives five candidates for $G'(r)$, and we find that 
only $G'=-2/((r^2+1)r)$
satisfies both $\Phi_3=0$ and $\Phi_5=0$.  The resulting conformal factor
is remarkably simple:
\be
\label{omega_burns}
\Omega=1+r^{-2},
\ee
and makes $\hat{g}$ scalar flat. To examine $\hat{g}$ set $r=1/\rho$, so that
\[
\hat{g}=g_B=d\rho^2+\frac{1}{4}\rho^2 (\sigma_1^2+\sigma_2^2+\sigma_3^2)+\frac{1}{4}(\sigma_1^2+\sigma_2^2)
\]
which is the Burns metric (\ref{burns_met}). We have
shown in Theorem \ref{theo_burns} that this metric does not admit
an isometric embedding in $\R^6$. This rules out the existence of the conformal
embedding of $\CP^2$ in $\R^6$, at least if $\Omega$ is only allowed to depend
on the radial coordinate.

The Burns metric 
isometrically embeds in $\R^7$. In the next section we shall show that 
the resulting  conformal embedding
of $\CP^2$ is equivalent to the standard one which results from the isometric embedding \cite{kobayashi} (see also \cite{HS} for a more recent application of 
this embedding) of $\CP^2$ in $S^7$.
\subsection{Canonical embedding of $\CP^2$ in $S^7$}
Let $H\cong \R^8$ be the 8 dimensional space of 3 by 3 Hermitian matrices with trace one. The pairing
\be
(A, B)=\mbox{Tr}(AB), \quad A, B\in H
\ee
induces a flat Euclidean metric\footnote{Alternatively, this flat metric is induced
on the hyper-plane $\mbox{Tr}(A)=1$ on the space $\R^9$ of all 3 by 3 Hermitian matrices.}  on $\R^8$, as $(A, A)= A_{ij}\ov{A}_{ij}$.

Let $Z=[Z^1, Z^2, Z^3]$ be homogeneous coordinates of a point in $\CP^2$.
A distance between two  points $Z, W\in \CP^2$ measured along geodesics of the Fubini--Study metric is
\[
d(Z, W)=2\arccos{\sqrt{\kappa(Z, W)}}, \quad \mbox{where}\quad \kappa(Z, W)=\frac{|<Z, W>|^2}{|Z|^2|W|^2}, \quad\mbox{and}\quad
<Z, W>\equiv \ov{Z}{}^\alpha{W}^\alpha.
\]
Using this notation, the Fubini--Study metric is the quadratic part of the expression
\[
\kappa(Z, Z+dZ)=1-\frac{|Z|^2|dZ|^2-|<Z, dZ>|^2}{|Z|^4}+O(dZ^3).
\]
From now on we shall impose a normalisation condition
$<Z, Z>\equiv \ov{Z}{}^\alpha{Z}^\alpha=1$.
Consider the embedding $\phi:\CP^2\rightarrow H$ given by
\be
\label{embeddigcp}
\phi(Z)=\left(\begin{array}{c c c}
  |Z^1|^2 & Z^1\ov{Z}{}^2 & Z^1\ov{Z}{}^3 \\ 
  Z^2\ov{Z}{}^1 & |Z^2|^2 & Z^2\ov{Z}{}^3\\
	Z^3\ov{Z}{}^1 & Z^3 \ov{Z}{}^2 & |Z^3|^2
 \end{array}\right).
\ee
This embedding is $SU(3)$ invariant, and isometric (in flat $\R^8$) as 
\[
\mbox{Tr}(\phi(Z) \phi(W))=|<W, Z>|^2.
\]
The normalisation  implies  $\mbox{Tr}(\phi(Z)^2)=1$, so this is also an isometric embedding of $\CP^2$ in $S^7$. Combining this result with the stereographic  projection form $S^7$ to $\R^7$ gives a conformal embedding of $\CP^2$ in $\R^7$, and
a natural question arises whether this conformal embedding is different
to the one we established in \S\ref{sectioncp2} via conformal equivalence with the Burns metric.
\begin{theo}
\label{theo52}
Let $\pi :S^7\rightarrow \R^7$ be the stereographic projection.
The map
\[
\pi\circ\phi:\CP^2\rightarrow \R^7
\]
is a conformal isommetric embedding of $\CP^2$, and the diagram
\begin{eqnarray}
(\CP^2, g_{FS}) &\overset{\phi}
{\longrightarrow}& S^7\subset \R^8\nonumber\\
\downarrow \Omega^2& & \downarrow \pi\;\\
\text{Burns metric}&\longrightarrow&\R^7\nonumber
\end{eqnarray}
is commutative. 
\end{theo}
\noindent
{\bf Proof.}
Consider a metric on $S^7$ induced from the flat $\R^8$
\be
\label{g7}
g_{S^7}=dR^2+R^2 g_{S^6}+dY^2=\frac{dR^2}{1-R^2}+R^2g_{S^6}, \quad\mbox{where}\quad  R^2+Y^2=1,
\ee
and $g_{S^6}$ is the round metric on the six--sphere. The metric $F^2 g_{S^7}$ with 
\be
\label{newcf}
F^2=\frac{1}{(1+\sqrt{1-R^2})^2}
\ee
is flat. 
To construct the embedding explicitly in the Bianchi--IX coordinates set
\be
\label{ZZ}
Z^1= \frac{r}{\sqrt{1+r^2}}\sin(\theta/2)e^{-i\phi/2}, \quad Z^2=\frac{r}{\sqrt{1+r^2}}\cos(\theta/2)e^{i\phi/2},
\quad Z^3= \frac{1}{\sqrt{1+r^2}}e^{-i\psi}.
\ee
The corresponding element of $H\cong \R^8$ is
\[
\phi(Z)=A=\left(\begin{array}{c c c}
  x & p & q \\ 
  \ov{p}& y & s\\
	\ov{q} & \ov{s}& 1-x-y,
 \end{array}\right)
\]
where  real coordinates $(x, y)$ and complex coordinates $(p, q, s)$ are now explicit functions of 
$(r, \phi, \theta, \psi)$
and we verify that $(d\phi(Z), d\phi(Z))$ is twice the Fubini--Study metric (\ref{cp2metric}).
The sphere equation 
$\mbox{Tr}(A^2)=1$ takes the form
\be
\label{sphere}
3(|p|^2+|q|^2+|s|^2)+\Big(\frac{3}{2}(x+y)-1\Big)^2+\Big(\frac{\sqrt{3}}{2}
(x-y)\Big)^2=1
\ee
and  (a constant rescaling of) the metric is
\be
\frac{3}{2}\mbox{Tr}(dA^2)=
|\sqrt{3}dp|^2+|\sqrt{3}dq|^2+|\sqrt{3}dr|^2+
d\Big(\frac{3}{2}(x+y)-1\Big)^2+d\Big(\frac{\sqrt{3}}{2}
(x-y)\Big)^2.
\ee
A choice of a coordinate $Y$  in (\ref{g7}) is equivalent to a choice of a point on $S^7$
from which the projection $\pi:S^7\rightarrow \R^7$ is made.
Every such choice
gives rise to a conformal factor  (\ref{newcf}). To reproduce
the Burns metric as the pull--back $(\pi \circ \phi)^* g_{\R^7}$ 
consider\footnote{We note that other choices of $Y$ give rise to non--radial conformal factors, and
conformal embeddings of $\CP^2$ in $S^7$.}
\[
Y=\frac{3}{2}(x+y)-1=\frac{3r^2}{2(1+r^2)}-1, \quad
\mbox{which gives}\quad F=\frac{1}{1+Y}=\frac{2}{3}(1+r^{-2})
\]
in agreement with  (\ref{omega_burns}) up to a constant
conformal factor. 
We now need to show that
not only the conformal factors, but also the embeddings agree in this case.
To do that, recall that a flat metric on $\R^8$
\[
\delta_{ij}dX^idX^j+dY^2, \quad i, j=1, \dots, 7
\]
gives rise to a metric on the sphere $\delta_{ij}X^iX^j+Y^2=1$ given by
\be
\label{stereo}
g_{S^7}=\frac{d{\xi_1}^2+\dots  +d{\xi_7}^2}{(1+{\xi_1}^2+\dots+{\xi_7}^2)^2},
\quad\mbox{where}\quad \xi_i=\frac{X^i}{1+Y}.
\ee
This is conformally flat, with the conformal factor 
$(1+Y)^{-2}$ as before. The flat Cartesian coordinates on $\R^7$ are given by
(\ref{stereo}) with \[
X^1+iX^2=\sqrt{3}Z^1\ov{Z}{}^2,\quad X^3+iX^4=\sqrt{3}Z^1\ov{Z}{}^3,\quad
X^5+iX^6=\sqrt{3}Z^2\ov{Z}{}^3,\quad X^7=\frac{\sqrt{3}}{2}(|Z^1|^2-|Z^2|^2).
\] 
This gives
\begin{eqnarray}
\xi_1+i\xi_2&=&\frac{1}{\sqrt{3}}\sin{\theta}e^{-i\phi}, \quad \xi_7=\frac{1}{\sqrt{3}}\cos{\theta},\\
\xi_3+i\xi_4&=&\frac{1}{\sqrt{3}}\rho \sin{(\theta/2)}e^{i(\psi-\phi/2)},\quad
\xi_5+i\xi_6=\frac{1}{\sqrt{3}}\rho\cos{(\theta/2)}e^{i(\psi+\phi/2)}\nonumber
\end{eqnarray}
where $\rho=r^{-1}$. The metric
$d{\xi_3}^2+d{\xi_4}^2+d{\xi_5}^2+d{\xi_6}^2$
is the flat $\R^4$ factor in (\ref{burns_met})
and $d{\xi_1}^2+d{\xi_2}^2+d{\xi_7}^2$ is the $S^2$ factor.
They add up to the Burns metric (\ref{burns_met}).\koniec
\section{Conformal embeddings of LRS Bianchi-type IX in $\R^7$}
\label{sectionLRS}
The name \emph{LRS Bianchi-type IX} means there is an isometry group locally isomorphic to $SU(2)\times U(1)$ transitive on 3-surfaces, with isotropy $U(1)$ at each point (so LRS stands for \emph{locally-rotationally symmetric}). Einstein or vacuum examples are, without loss of generality, diagonal in a basis of invariant 1-forms \cite{todcoh} so we consider only the metric form:
Let
\be
\label{b9l}
g=adr^2+b(\sigma_1^2+\sigma_2^2)+c\sigma_3^2, 
\ee
and let
\be
\label{b92}
\eta=dR^2+\frac{1}{4}R^2(\sigma_1^2+\sigma_2^2+\sigma_3^2)+
\epsilon(dZ^2+Z^2(\sigma_1^2+\sigma_2^2))
\ee
be a pull back of a flat metric from $\R^7$ (if $\epsilon=1)$
or $\R^{4, 3}$ (if $\epsilon=-1$) to a five--manifold obtained by
identifying the angles of the two--sphere factors in $\R^4$ and $\R^3$. 
The radial conformal embedding condition
\be
\label{formulae_rz}
\Omega^2g=\eta, \quad \Omega=\Omega(r), \quad R=R(r),\quad  Z=Z(r)
\ee
gives
\[
R^2=4\Omega^2c, \quad Z^2=\epsilon\Omega^2(b-c)\geq 0 \quad\mbox{(this determines the sign of 
$\epsilon$)}
\]
and
\be
\label{omega}
\Omega=\exp{\Big(\int\frac{-\dot{c}-\epsilon\dot{h}\pm\sqrt{a(c+\epsilon h
)-\frac{(c\dot{h}-h\dot{c})^2}{\epsilon ch} }}{2(c+\epsilon h)}dr\Big)}, 
\quad\mbox{where}\quad
h=\frac{1}{4}(b-c).
\ee
Therefore in general there are two conformal factors which make the embedding possible.
They are related by the involution
\[
\Omega\rightarrow\frac{1}{(c+\epsilon h)\Omega}.
\]
Now look at some ASD examples which will all have $\epsilon=1$. 
We know (Corollary 2.4 in \cite{DT19})
that all these must have a conformal embedding class at least $2$. 
The calculations below show that for each of these examples the conformal embedding class is at most $3$

\begin{itemize}
\item The Burns metric has 
\[
a=1, \quad b=\frac{1}{4}(r^2+1), \quad c=\frac{1}{4}r^2,
\]
which gives 
\[
\Omega=1, \quad\mbox{or}\quad\Omega=\frac{1}{1+4r^2}
\]
depending on the choice of sign in the integral (\ref{omega}).
This embedding is global.
\item The Eguchi--Hanson metric \cite{EHref} has
\[
a=\Big(1-\frac{a^4}{r^4}\Big)^{-1}, \quad b=\frac{1}{4}r^2, \quad
c=\frac{1}{4}r^2\Big(1-\frac{a^4}{r^4}\Big),\quad\mbox{where}\quad r>a
\]
which gives
\[
\Omega=\exp{\Big(\int\frac{3a^4+4r^4\pm 2a^4
\sqrt{\frac{7a^4-4r^4}{a^4-r^4}}}{r(3a^4-4r^4)}dr\Big)}.
\]
This only covers the $r>(7/4)^{1/4}a$ range of the EH--manifold,
and even in this range the embedding is only local, as the regularity
of the Eguchi--Hanson metric at $r=a$ requires the range of $\psi$ to
be $0\leq \psi\leq 2\pi$ (rather than $0\leq \psi\leq 4\pi$
which is used in to cover $\R^7$ in (\ref{b92})).
A global isometric embedding of the Eguchi--Hanson manifold in
$\R^{11}$ has been presented in \cite{EHembedding}.
\item The Fubini--Study is conformal to Burns. It has
\[
a=\frac{1}{(r^2+1)^2}, \quad b=\frac{1}{4}\frac{r^2}{r^2+1}, \quad
c=\frac{1}{4}\frac{r^2}{(r^2+1)^2}
\]
and we find
\[
\Omega=1+r^{-2},\quad\mbox{or}\quad \Omega=\frac{r^2+1}{r^2+4},
\]
the first of which we already knew from (\ref{omega_burns}).
\item The Taub--NUT metric has
\[
a=\frac{1}{4}\frac{r+m}{r-m}, \quad b=\frac{1}{4}(r^2-m^2), \quad
c=m^2\frac{r-m}{r+m}, \quad r>m
\]
which gives
\[ 
\Omega=\exp{\Big(\int\frac{(3m+r)(4m^2-mr+r^2)\pm(r+m)^2\sqrt{\frac{55m^3+3m^2r+5mr^2+r^3}{3m+r}}  }{(m^2-r^2)
(13m^2+2mr+r^2)}dr\Big)}.
\]
This covers the whole range $r>m$, and MAPLE computes $\Omega$ in terms of elliptic 
integrals.
To analyse the apparent singularity at $r = m$ set $r = m +\rho^2/(2m)$ so
that near $r = m$ the metric is flat
\[
g\sim d\rho^2+\frac{\rho^2}{4}\Big(\sigma_1^2+\sigma_2^2+\sigma_3^2 \Big).
\]
Making the same coordinare change in $\Omega$ yelds two conformal factors: 
$\Omega_1\sim 1$, and $\Omega_2\sim\rho^{-1}$ near $\rho = 0$. Using 
$\Omega_1$ for conformal rescalling gives a 
flat metric near the
NUT point $r = m$. The second conformal factor moves the NUT to $\infty$, which can
be seen by replacing making a coordinate transformation $\rho=1/\hat{\rho}$ in 
$\hat{g} = {\Omega_2}^2g$. The
resulting metric is also flat near $\hat{\rho}= 0$.
Therefore the resulting conformal embedding is global.
\item The Agaoka obstructions (\ref{a2}) do not vanish for the ASD Taub--NUT and the Eguchi--Hanson metrics.
Moreover, there does not exist a conformal factor depending only on $r$ and such that conformal rescallings 
of Eguchi--Hanson or ASD  Taub NUT have vanishing (\ref{a2}). In fact  a MAPLE aided computation shows
that the only non--conformally flat LRS Bianchi IX metric with ASD conformal curvature and vanishing Agaoka's 
obstructions is the Burns metric.
\end{itemize}
\subsection{Isometric embeddings in $\R^7$.}
In \S\ref{sectionLRS} we listed a number of examples of conformally anti--self--dual
metrics of the form (\ref{b9l}) which admit a radial conformal embedding in $\R^7$.
Of  these only the Burns metric (\ref{burns_met}) admits an embedding with $\Omega=1$, that is an  isometric 
embedding. We shall now show that under the additional assumption of zero scalar curvature
the Burns metric is the only metric with this property
\begin{theo}
\label{theounib}
Let $(M, g)$ be a non--flat LRS Bianchi IX Riemannian manifold with anti--self--dual conformal curvature
and zero Ricci scalar, and let $\iota: M\rightarrow \R^7$ be a radial isometric embedding
(\ref{b92}), (\ref{formulae_rz}) such that $\iota^*(\eta)=g$. Then $g$ is isometric to a constant multiple
of the Burns metric (\ref{burns_met}).
\end{theo}
\noindent
{\bf Proof.}
First recall \cite{todp6} that any diagonal Bianchi IX metric can be put in the form
\be
\label{paul_metric}
g=w_1w_2w_3 dt^2+\frac{w_2w_3}{w_1}{\sigma_1}^2+ \frac{w_1w_3}{w_2}{\sigma_2}^2
+\frac{w_1w_2}{w_3}{\sigma_3}^2
\ee
for some $w_j=w_j(t)$ where $j=1, 2, 3$. Define three more functions
$a_j=a_j(t)$ by
\be
\label{paula1}
\dot{w}_1=-w_2w_3+w_1(a_2+a_3) \quad \mbox{(and cyclic permutations)}.
\ee
Then the scalar--flat anti--self--duality equations are
\be
\label{paulwa}
\dot{a}_1=-a_2a_3+a_1(a_2+a_3) \quad \mbox{(and cyclic permutations)}.
\ee
We restrict to the LRS Bianchi IX metrics (\ref{b9l}), and
set 
\[
w_1=w_2=w, \quad w_3=u, \quad a_1=a_2=a, \quad a_3=A.
\]
The equations (\ref{paulwa}) and (\ref{paula1}) simplify to
\begin{subequations}
\begin{align}
\dot{w}&=-uw+(a+A)w\label{g1}\\
\dot{u}&=-w^2+2au\label{g2}\\
\dot{a}&=a^2\label{g3}\\
\dot{A}&=-a^2+2aA.\label{g4}
\end{align}
\end{subequations}
Comparing expressions (\ref{b92}) with (\ref{paul_metric}) we find
\[
R^2=4\frac{w^2}{u}, \quad Z^2=u-\frac{w^2}{u},
\]
when now $R$ and $Z$ are regarded as functions of $t$.

The isometric embedding condition  $\dot{R}^2+\dot{Z}^2=w^2u$ takes the form
\be
\label{burns_emb}
\Big(2\dot{w}-w\frac{\dot{u}}{u}\Big)^2+        
\frac{\Big(\dot{u}-\frac{2 w\dot{w}}{u}+ \frac{w^2\dot{u}}{u^2}\Big)^2}{4(u^2-w^2)}=u^2w^2.
\ee
Equation (\ref{g3}) is the Riccati equation with the general solution $a=(t_0-t)^{-1}$, and a singular integral $a=0$. The constant
$t_0$ can be set to $0$ by a translation of the $t$--coordinate, and the remaining equations in (\ref{g1}--\ref{g4})  can be readily solved:
\subsubsection*{The singular case:}
The coupled system of equations (\ref{g1}--\ref{g4}) can be solved to give
\be
a=0,\quad A=c, \quad w=\frac{4 e^{t/c_1}}{e^{2t/c_1}{c_1}^2-4}, 
\quad u=\frac{e^{2t/c_1}{c_1}^2(c_1c+1)+4(1-c_1c)}{c_1(e^{2t/c_1}{c_1}^2-4)}
\ee
where one of the integration constants has been elliminated  by translating the
$t$--coordinate, and $c, c_1$ are the remaining constants of integration.
The embedding condition (\ref{burns_emb}) forces $c_1=\pm c^{-1}$, and w. l. g.
we can choose $c_1=c^{-1}$ by changing the sign of $t$ if neccessary. This gives
\[
u=\frac{2ce^{2tc}}{e^{trc}-4c^2}, \quad w=2c e^{-tc}u,
\]
with the range of $c$ now restricted by $0<c<1/2$ for regularity.

We claim that the resulting metric is a constant rescalling of the Burns 
solution (\ref{burns_met}). To see this, define $\rho$ by
\[
\frac{w^2}{u}=\frac{1}{4}\rho^2, 
\]
so that
\[
t=\frac{1}{2c}\ln{\Big( \frac{4c^2(\rho^2+8c)}{\rho^2}\Big)}, 
\quad u=\frac{\rho^2}{4}+2c, \quad w=\frac{\rho}{4}\sqrt{\rho^2+8c}.
\]
We verify that $uw^2dt^2=d\rho^2$, so that finally
\be
\label{almost_b}
g=d\rho^2+\frac{\rho^2}{4}({\sigma_1}^2+{\sigma_2}^2+{\sigma_3}^2)
+2c({\sigma_1}^2+{\sigma_2}^2)
\ee
which agrees with the Burns metric (\ref{burns_met}) if $c=1/8$. 
For any other non--zero value of $c$,
the coordinate transformation $\rho\rightarrow \sqrt{8c}\rho$ gives
$g$ as  $8c$ times the Burns metric. If $c=0$ then the metric
(\ref{almost_b}) is flat.
\subsubsection*{The generic case}
Setting $t_0=0$ in the general solution of the Riccati equation
(\ref{g3}), and solving the remaining equations in (\ref{g1}--\ref{g4}) gives
\be
a=-\frac{1}{t}, \;\;\; A=-\frac{1}{t}+\frac{c}{t^2},\quad 
w=\frac{4e^{\frac{c_2t+1}{c_1}}}
{t^2\Big({c_1}^2e^{\frac{2}{c_1t}} -
4 e^{\frac{2c_2}{c_1}}\Big)}, \quad u=-\frac{\dot{w}}{w}+\frac{c}{r^2}-\frac{2}{r}.
\ee
The embedding condition (\ref{burns_emb}) does not hold for any values
of the integration constants $c, c_1, c_2$, which can be seen by looking at
the coefficients of various powers of $\gamma\equiv\exp{(-1/(c_1t))}$ in (\ref{burns_emb}). The coefficient of $\gamma^{20}$ vanishes iff $c_1=1/c$, but then
the coefficent of $\gamma^2$ is non--zero.
\subsection{LRS Bianchi-IX isometrically embedded into flat $\mathbb{R}^8$}
\label{sub71}
This can be done with signatures depending on the example considered. Consider the flat metrics
\[dR^2+\frac{R^2}{4}(\sigma_1^2+\sigma_2^2+\sigma_3^2)+\epsilon_1(dZ^2+Z^2(\tilde{\sigma}_1^2+\tilde{\sigma}_2^2))+\epsilon_2dF^2,\]
with $\epsilon_i=\pm 1$ and embedding
\[\tilde\theta=\theta,\;\;\tilde\phi=\phi,\;\;R(t),\;\;F(t),\;\;Z(t).\]
This gives (\ref{b9l}) if we choose
\[R^2=4c,\;\epsilon_1Z^2=b-c,\;\;{(R')}^2+\epsilon_1 {(Z')}^2+\epsilon_2{(F')}^2=a.\]
The choices of $\epsilon_1,\epsilon_2$ are dictated by the requirement of reality for $F$ and $Z$. 
\subsubsection{$\mathbb{CP}^2$}
With $\epsilon_1=\epsilon_2=1$ set
\[R=\frac{r}{1+r^2},\;Z=\frac{r^2}{2(1+r^2)},\;F=\frac{\sqrt{3}}{2(1+r^2)},\]
then the metric is Fubini-Study. Since for this embedding we have
\[2R^2+4Z^2+4F^2/3=1,\]
some rescaling of the flat coordinates gives this as the familiar embedding of $\mathbb{CP}^2$ into ${S}^7$.
\subsubsection{Eguchi-Hanson}
With $\epsilon_1=1=-\epsilon_2$ set
\[R=\frac{(r^4-a^4)^{1/2}}{r},\;Z=\frac{a^2}{2r}\]
and solve
\[F'=\frac{\sqrt{3}a^2}{2r^2}\left(\frac{3r^4+a^4}{r^4-a^4}\right)^{1/2},\]
for $F(r)$ to obtain the Eguchi-Hanson metric.
\subsubsection{Anti--self--dual Taub-NUT}
With $\epsilon_1=\epsilon_2=1$ set
\[R=2m\left(\frac{r-m}{r+m}\right)^{1/2},\;Z=\frac{(r-m)}{2}\left(\frac{r+3m}{r+m}\right)^{1/2}\]
and solve
\[F'=\left(\frac{2mr^3+9m^2r^2+24m^3r+29m^4}{4(r+m)^3(r+3m)}\right)^{1/2}\]
to obtain the ASD Taub-NUT metric.

\end{document}